# Band energy diagrams of n-GaInP/n-AlInP(100) surfaces and heterointerfaces studied by X-ray photoelectron spectroscopy

Mohammad Amin Zare Pour[1], Oleksandr Romanyuk[2], Dominik C. Moritz[3], Agnieszka Paszuk[1], Clément Maheu[3], Sahar Shekarabi[1], Kai Daniel Hanke[1], David Ostheimer[1], Thomas Mayer[3], Jan P. Hofmann[3], Wolfram Jaegermann[3], Thomas Hannappel[1,*]

[1]*Institute of Physics, Fundamentals of Energy Materials, Technische Universität Ilmenau, Gustav-Kirchhoff-Str. 5, 98693 Ilmenau, Germany*

[2]*FZU – Institute of Physics of the Czech Academy of Sciences, Cukrovarnicka 10, 16200 Prague, Czech Republic*

[3]*Surface Science Laboratory, Department of Materials and Earth Sciences, Technical University of Darmstadt, Otto-Berndt-Strasse 3, 64287 Darmstadt, Germany*

## Abstract

Lattice matched n-type AlInP(100) charge selective contacts are commonly grown on n-p GaInP(100) top absorbers in high-efficiency III-V multijunction solar or photoelectrochemical cells. The cell performance can be greatly limited by the electron selectivity and valance band offset at this heterointerface. Understanding of the atomic and electronic properties of the GaInP/AlInP heterointerface is crucial for the reduction of photocurrent losses in III-V multijunction devices. In our paper, we investigated chemical composition and electronic properties of n-GaInP/n-AlInP heterostructures by X-ray photoelectron spectroscopy (XPS). To mimic an in-situ interface experiment with in-situ stepwise deposition of the contact material, 1 nm – 50 nm thick n-AlInP(100) epitaxial layers were grown on n-GaInP(100) buffer layer on n-GaAs(100) substrates by metal organic vapor phase epitaxy. We observed (2×2)/c(4×2) low-energy electron diffraction patterns with characteristic diffuse streaks along the $[01\bar{1}]$ direction due to P-P dimers on both AlInP(100) and GaInP(100) as-prepared surfaces. Atomic composition analysis confirmed P-rich termination on both surfaces. Angle-resolved XPS measurements revealed a surface core level shift of 0.9 eV in P 2p peaks and the absence of interface core level shifts. We assigned the surface chemical shift in the P2p spectrum to P-P bonds on a surface. We found an upward surface band bending on the (2×2)/c(4×2) surfaces most probably caused by localized mid-gap electronic states. Pinning of the Fermi level by localized electronic states remained in n-GaInP/n-AlInP heterostructures. A valence band offset of 0.2 eV was derived by XPS and band alignment diagram models for the n-n junctions were suggested.

## 1. Introduction

Ternary III-V semiconductors are widely used as a promising photoabsorber material in photoelectronic or photoelectrochemical devices including light-emitting diodes, photodetectors, electrooptic modulators, and frequency-mixing components due to their tunable band gaps and tunable band alignment at the heterojunction. In particular, the application of ternary III-V photoabsorbers for solar cells or photoelectrochemical devices enables an efficient absorption of the solar spectrum [1] allowing to reach word record solar-to-electricity [2] and solar-to-hydrogen [3,4] efficiencies.

GaInP has an ideal band-gap engineering potential and is commonly grown as a top photoabsorber with either lattice matched stoichiometry ($Ga_{0.51}In_{0.49}P$) to GaAs in highly-efficient 3-junction solar cells on Ge(100) [5] or on Si(100) substrates with world-record efficiencies [6]. It can be also grown lattice-matched to an underlying GaInAs bottom solar cell in a 2-junction solar [7] and world-record efficiency photoelectrochemical devices [3–6]. AlGaInP quinternary compound with an Al-content of 18% (2.1 eV) have been employed as the top photoabsorber in six-junction solar cells for the highest direct bandgap among III-V semiconductors lattice-matched to GaAs [2]. The ternary, $Al_xIn_{1-x}P$ material is currently the most preferred for the optically transparent so called 'window' layer for electron collection in 2- or 3-junction solar cells or photoelectrochemical devices. This charge selective contact should passivate the surface, conduct the majority carriers and reflect the minority carriers [8]. In this regard, the influence of the band structure at the heterointerface on solar cell performance was investigated for GaAs/GaInP and for GaAs/AlInP [9,10]. The performance of such devices can be strongly limited by recombination losses at the heterointerface. Fermi level pinning at the surface of the AlInP window layer can result in a large increase in interface recombination velocity due to surface states [11], which can be reduced through passivation or functionalization of the surface towards $PO_x$ as shown for photoelectrochemical cells [4]. Therefore, the investigation of the electronic structure at the critical III-V heterointerfaces is essential for designing highly efficient devices, as defects can inhibit charge transfer, and the band alignment affects the performance of the device. The effect of bands offset plays a role in determining the materials limits, and also the electronic properties [12].

---

[*] Corresponding author's email: thomas.hannappel@tu-ilmenau.de

Metalorganic vapor phase epitaxy (MOVPE) allows for the preparation of III-V semiconductor photoabsorbers at industrial scale production i.e., with extremely high growth rates and high precursor incorporation efficiencies, with a very good crystalline quality employing wafers with large size and high quantity in a single process [13,14].

In complex multilayer solar or photoelectrochemical device structures, the III-V heterointerfaces must be abrupt for efficient carrier transport, thus, a well-ordered surface prior to growth of a subsequent layer must be prepared. Interfacial sharpness is in particular crucial for the interface between the top cell and the charge selective contact. Depending on the P chemical potential in the MOVPE reactor with $H_2$ carrier gas, the GaInP(100) surface can be terminated by either phosphorus dimers, so called 'P-rich' or by group-III-P dimers, so called 'group-III-rich' surface. In analogy to GaP(100) and InP(100), the P-rich surface exhibits $(2\times2)/c(4\times2)$ surface reconstruction and the group-III-rich surface, $(2\times4)$ surface reconstruction. These surface reconstructions were investigated quite extensively by theory and experiment [15–21]. So far, only few studies focused on the surface reconstruction of AlInP(100) epilayers [22,23].

Photoelectron spectroscopy is a widely used technique to gain quantitative information about atomic composition and electronic properties of the materials. The information depth of PES is governed by the photoelectron inelastic mean free path (IMFP) [24]. Laboratory based X-ray sources such as Al Kα (1486.74 eV) enable analysis of surfaces or near-surface regions (<10 nm). Higher IMFPs of photoelectrons and deeper probe depth require higher excitation energies, which are typically available at synchrotrons, i.e., so-called hard X-ray photoelectron spectroscopy (HAXPES) [25]. In the presence of strong band bending in heterointerfaces, the measured values of valence band offsets (VBO) may deviate from the band discontinuity exactly at interface [26]. The band alignment and the electronic states at the GaInP/AlInP(001) interfaces have been investigated theoretically [27,28] while there is a lack of experimental confirmation about band offsets and band alignment at this heterojunctions. In this paper, we utilized soft-X-ray photoelectron spectroscopy (XPS) for the measurement of the VBOs [29] in the thin GaInP/AlInP heterostructures.

## 2. Experimental part

In order to mimic an in-situ interface experiment with stepwise deposition of the contact material, different samples with different thickness of AlInP(100) layers (between 1 nm – 50 nm) were grown in a $H_2$-based, horizontal-flow MOVPE reactor (Aixtron, AIX-200) on n-type GaInP/GaAs (100) wafer (Si-doped, $2 \times 10^{18}$ atoms/cm$^3$) with 0.1° offcut toward <111> direction. Prior to AlInP growth, the GaAs(100) substrate was thermally deoxidized at 620°C under tertiarybutylarsine [30] flow. Subsequently, 100 nm thick GaAs(100) buffer layer and 100 nm thick GaInP(100) buffer layer were grown at 600 °C and pressure of 100 mbar using tertiarybutylphosphine (TBP), trimethylindium (TMIn), trimethylgallium (TMGa), trimethylaluminum (TMAl). The ditertiarybutyl silane (DTBSi) was used as the n-type dopant source for all grown buffers. The molar V/III ratio during GaInP and AlInP growth was set to 61 and 63, respectively (P-rich growth conditions). The AlInP(100) growth rate of 0.3 nm/s was obtained on reference thick samples by *in situ* reflection anisotropy spectroscopy (RAS, LeyTec) measurements of the period of Fabry-Perot oscillations [31].

After growth, the sample surfaces were cooled down to 300 °C in presence of the TBP precursor to avoid desorption of P from the surface at elevated temperatures. Subsequently, the TBP source was closed, and the samples were annealed at 310 °C for 10 min to remove precursor residuals and excess P from the surface [32].

The as-prepared samples were transferred from the MOVPE reactor to the surface-sensitive analytical tools via a transfer shuttle under ultra-high vacuum (UHV) conditions [33]. The UHV cluster is equipped with low-energy electron diffraction (LEED, Specs ErLEED 100-A) and XPS (Specs Focus 500/Phoibos 150/1D-DLD-43-100, monochromated Al-Kα line). Spectrometer work function was calibrated by using the Au $4f_{7/2}$ core level at 84.00 eV measured on a sputtered Au reference sample. Angle-resolved (AR) XPS measurement were carried out at normal emission geometry (90°) and at sample tilted geometry (30°, surface sensitive). The high-resolution core level peaks were measured with pass energy of 10 eV, energy step of 0.1 eV, and energy resolution of 0.6 eV (verified by measuring the full width at half maxima (FWHM) of Ag $3d_{5/2}$ core level). The high-resolution core levels were fitted with Voigt function (Gaussian-Lorentzian mixing ratio of 1.7-1.9 for all the core levels) profiles after subtraction of a Shirley background. The chemical composition of the film was determined from the corresponding fitted peak areas by using the relative sensitivity factors [34].

The carrier concentration depth profile in the III-V epilayers was measured *ex situ* by electrochemical capacitance voltage profiling (ECVP, WEP-CVP 21) with 0.1 M HCl solution (under visible light illumination at room temperature). The molar flow of DTBSi was adjusted so that the carrier concentration in the n-GaAs homoepitaxial buffer layer is ~$4\times10^{18}$ cm$^{-3}$, in the n-GaInP layer ~ $1\times10^{17}$ cm$^{-3}$ and n-AlInP ~$1\times10^{17}$ cm$^{-3}$.

The lattice constants of the grown overlayers were measured *ex situ* by high-resolution X-ray diffractometry (HR-XRD) ω/2θ scans (Bruker AXS D8 Discover with Ge(022)x4 asymmetric monochromator and Goebel mirror). We verified that overlayer lattice constants were matched to the GaAs(100) substrate lattice constant (see SI, Fig. SI1). XRD measurements of atomic stoichiometry in the bulk confirms x=0.52 and y=0.51 in $Al_xIn_{1-x}P$ and $Ga_yIn_{1-y}P$, respectively. The present composition corresponds to lattice constant of a=5.653 Å and band gap of 2.3 eV (AlInP) [35] and 1.8 eV (GaInP) [36].

## 3. Results and discussion

### 3.1 Surface reconstruction

Fig. 1 shows LEED patterns of a) GaInP(100), b) 3-nm-thick GaInP/AlInP(100), and c) 11-nm-thick GaInP/AlInP(100) surfaces measured with primary electron beam energies between 51 eV - 68 eV. A (1×1) unit cell is indicated by dashed rectangular. All LEED patterns show half-order spots along the [011] direction and diffuse half-order streaks along the [0$\bar{1}$1] direction, which indicates a 2×1-like surface reconstruction. Similar LEED patterns were previously observed on P-rich GaP(100) and InP(100) surfaces consisting of rows of P-P buckled dimers partially saturated by hydrogen (one hydrogen atom saturates one P dangling bond) [19,37,38]. Diffuse streaks are formed due to random occupation of H within a surface unit cell, i.e. superposition of unit cells with different translation periodicities such as (2×2)/c(2×4). A diversity of unit cells causes one-dimensional disorder [39] and, as result, diffuse streaks on LEED patterns. The similar LEED patterns were observed on all GaInP/AlInP samples with different overlayer thickness. For thicker AlInP films, however, the diffraction spots became more diffuse, and the background intensity increases. Therefore, LEED data suggest that AlInP(100) surfaces are less ordered than the surface of the GaInP(100) layer. Recent theoretical studies of AlInP(100) surface reconstruction confirm stability of buckled P-P dimers with one H atom per dimer under MOVPE experimental conditions [40].

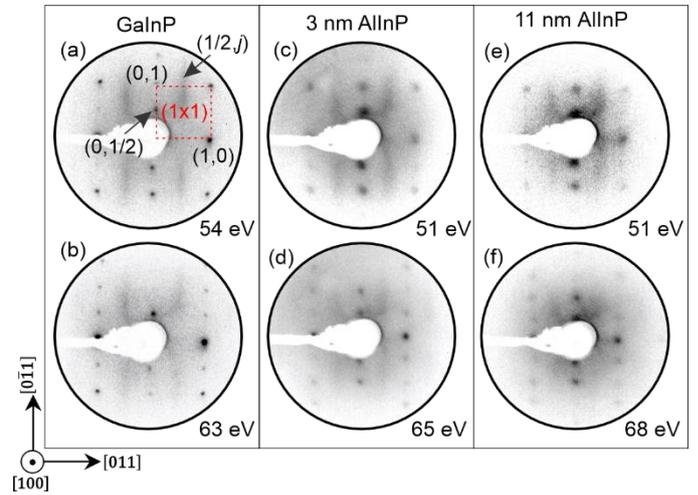

Fig. 1, LEED patterns of P-rich a), b) GaInP(100), c), d) 3-nm-thick GaInP/AlInP(100), and e), f) 11-nm-thick GaInP/AlInP(100) sample surfaces. A (1x1) unit cell is indicated by dashed red rectangular. Fractional-order spots indicates P-rich (2x2)/c(4x2) surface reconstruction.

Fig. 2 shows a) Al 2p, b) In $3d_{5/2}$, c) P 2p, d) Ga $2p_{3/2}$ core level spectra and e) valence band spectra measured XPS on GaInP sample (0 nm) and GaInP/AlInP samples (1 nm - 50 nm). All peak intensities were normalized for better clarity. Measured binding energies are included in Tab. SI2 in SI. In Fig. 2 d) a neglectable signal of Ga (<0.1 at. %) was still present above the limit of XPS information depth (~10 nm). The Ga residuals on the surfaces originates from desorption of Ga form the MOVPE reactor's walls. No O nor C was detected on all surfaces.

The core level peak positions and the valence band maxima (VBM) depend on overlayer thickness. Relative energy shifts with respect to position on 0 nm sample (GaInP(100) surface) and 1 nm sample for Al 2p is shown in Fig. 2 f). The mean standard deviations are ±0.05 eV (the same as the energy steps) for all the core levels and ±0.1 eV for the VBM. The core level peaks and VBMs shift toward the higher binding energy with increase of the AlInP thickness up to 11 nm. Shifts turned sign toward lower binding energy for larger thicknesses. Note, up to 11 nm, shift directions are similar for all core levels and VBM but shift magnitudes are different. After 11 nm, all the core levels and VBM shift to lower binding energy to the same extent.

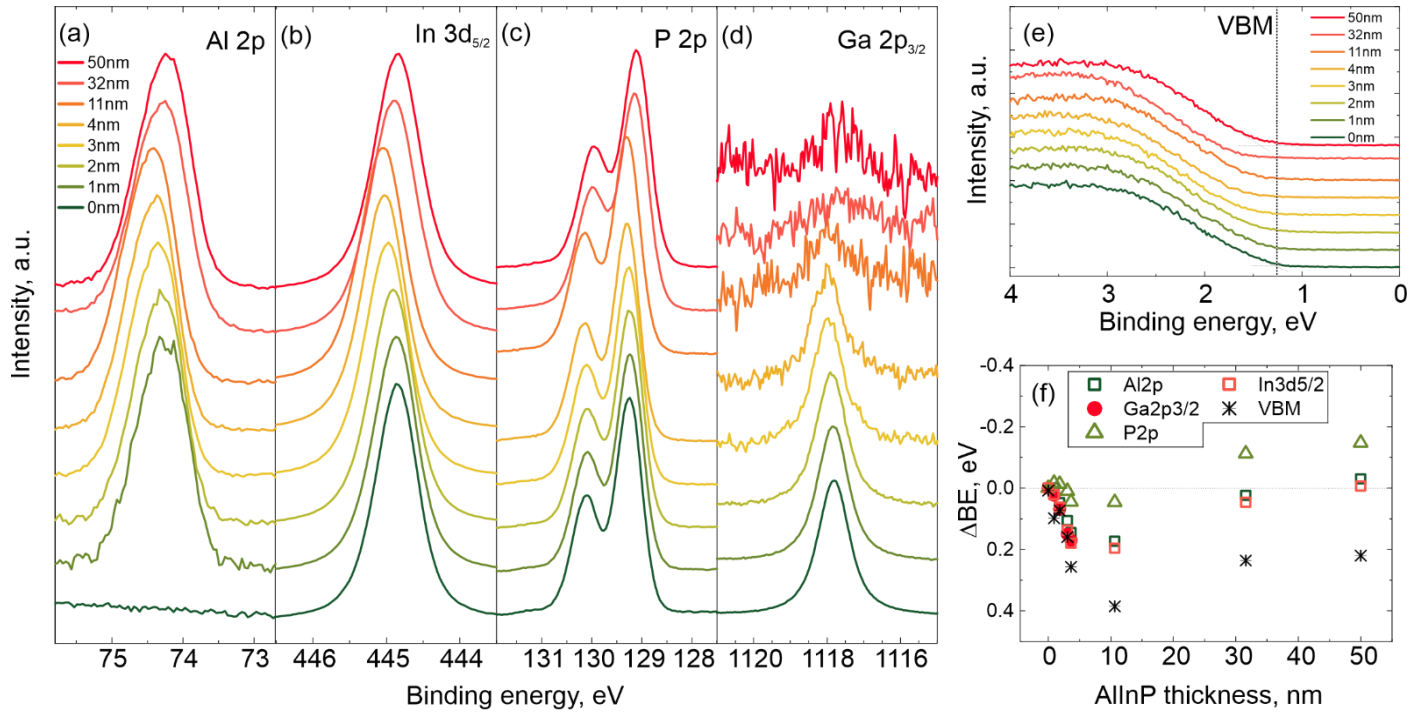

Fig. 2, a-d) XPS core level and e) VB spectra dependence on the thickness of AlInP overlayer. f) Relative binding energy shifts in respect with the values of 0 nm sample. Shifts area related to the chemical composition variation at initial stages of growth and to the surface band bending.

The origin of the core level shifts could be related to surface or interface band bending (BB) as well as due to the change of the atomic stoichiometry at interface (chemical shifts). In case of BB, one should expect almost similar shift magnitudes for the core levels with similar binding energies. However, this is not the case here - shift magnitudes are different. Therefore, apart from BB change (due to small changes in doping concentration or surface charge change), core level shifts due to changes in the chemical environment of atoms in lattice (which could be the element specific) should be also present.

The measured VBM values lie between 1.2 eV (0 nm) and 1.4 eV (50 nm sample) with respect to the Fermi level (zero eV). Based on ECVP measurements (see Fig. SI2 in SI), the estimated Fermi level position should be around 0.05 eV below the conduction band minimum (CBM) [41], i.e., around 1.75 eV and 2.25 eV for GaInP and AlInP above the VBM, respectively. Lower VBMs could be explained by the upward surface BB caused by surface charges due to presence of surface or defect electronic states. Therefore, the Fermi level position is pinned in the band gap of semiconductors. The overall experimentally observed VBM shift is therefore due to band bending as well as the change in the bandgap from GaInP to AlInP, i.e., due to valence band offset.

Fig. 3 a) shows the schematic experimental setup for the AR-XPS measurements. In normal emission geometry, the information depth is larger than in the tilted geometry (more surface sensitive conditions). AR-XPS spectra are shown for the 3-nm-thick sample [Fig. 3 b)-e)] and 11-nm-thick sample [Fig. 3 f)-i)]. Former dataset includes intensity contributions from both substrate and overlayer, i.e., including interface and surface; the latter dataset involves intensity from the AlInP overlayer, i.e., including surface only.

The P 2p core level peaks are fitted by two doublets consisting of two components representing $2p_{3/2}$ and $2p_{1/2}$ spin orbit splitting with area ratio of 1:2. The largest (green color) doublet represent intensity contribution from the bulk-like bonds (III-P bonds). The area of the second doublet (purple color) increases with emission angle decrease. Therefore, this component is associated with P bonds on a surface. The shift of the second doublet is 0.9 eV in respect to the bulk component. Positive shift in P 2p peak is related to P-P bonds [42], which in our case, should be P-P dimer bonds on the (2×2)/c(4×2) reconstructed surface. Two doublets in P 2p peak were also resolved on 11-nm-thick sample surface [Fig. 3 i)] (without interface contribution). Therefore, no interface components present in the P 2p core levels.

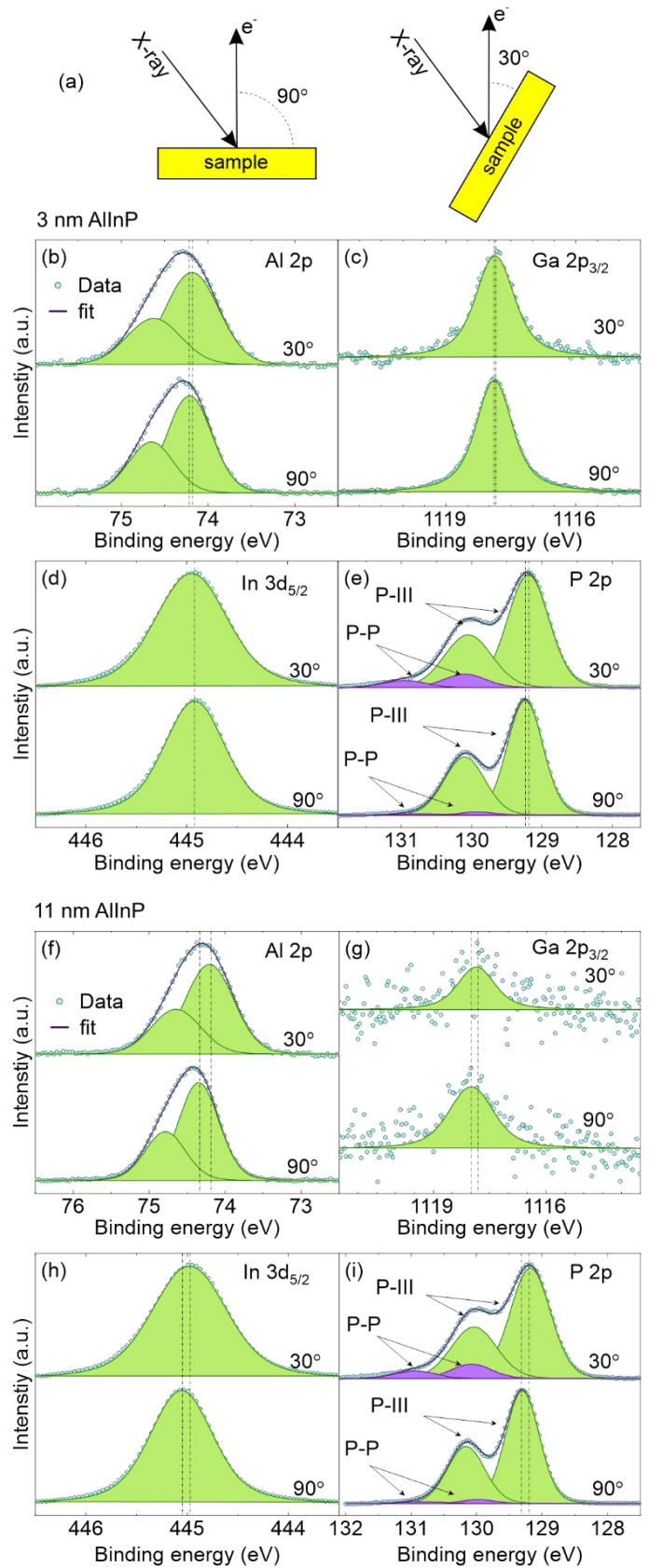

Fig. 3, a) Schematic plot of AR-XPS experimental setup. b)-e) Core level spectra measured on 3-nm-thick GaInP/AlInP(100) sample and f)-i) on 11-nm-thick sample. Bulk and surface components are indicated by green and purple color, respectively.

No surface or interface components were, detected in the Al 2p, Ga $2p_{3/2}$, and In $3d_{5/2}$ core levels (the Al 2p peak consist of one bulk doublet, see Fig. 3). Therefore, III-P bonds at the buried heterointerface should be for atoms bulk-like despite some surplus P atoms situated at the surface (see Fig. 4 and its discussion below). The crystal lattice type should be preserved across the interface with only cation atom type variation in the zinc blende lattice. This indicates formation of abrupt interface structure.

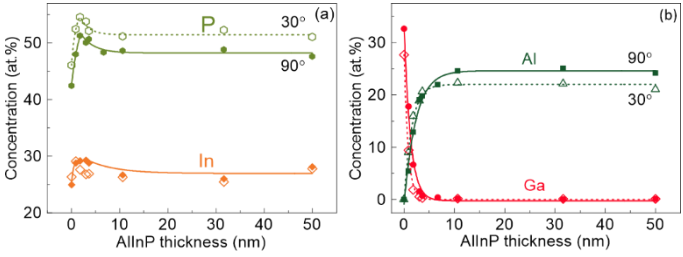

Fig. 4, Atomic concentration of a) P, In, and b) Al and Ga measured by AR-XPS on GaInP/AlInP(100) samples with different overlayer thicknesses. P-rich surfaces were confirmed for both GaInP and AlInP samples.

Dashed vertical guidelines in Fig. 4 show the change of the fitted peak maxima with the change of the emission angle. Shift to the lower binding energy with decreasing emission angle corresponds to the upward surface BB. Therefore, the AR-XPS measurements of the direction of the BB agree with the VBM measurements in Fig. 2 e). Shifts are more obvious for the 11 nm sample than for the 3 nm sample, where the variation of atomic composition may shift the core levels independently from the BB. In order to verify the effect of the chemical shifts, we compared the relative energy differences between the core level pairs in the GaInP/AlInP heterostructures with respect to the corresponding differences of the bulk references (see Fig. SI3, Fig. SI4, and Tab. SI1 in SI). Thin AlInP overlayers <10 nm show variation of core level positions distances, which indicates different stoichiometry of each overlayer.

The atomic composition of the samples was derived from the integrated core level peak intensities weighed with the respective sensitivity factors. In Fig. 4, atomic concentration dependence on AlInP overlayer thickness is shown. Close (open) circles correspond to data obtained using normal (tilted) sample geometry. In Fig. 4 a), P and In atomic composition varies with overlayer thickness up to ~10 nm and saturates for thicker films. In Fig. 4 b), concentration of Al (Ga) increases (decreases) up to ~10 nm the intensity from the GaInP substrate decreased with overlayer thickness increase due to damping by the increasing AlInP overlayer thickness. It seems to be nicely exponential, which would indicate layer by layer growth.

P atomic concentration increases on a surface (see 30° data points in Fig. 4 a). Surfaces of both GaInP and AlInP are the P-rich surfaces. Al layer lies below the P topmost layer according to AR-XPS measurements in Fig. 4 b).

### 3.2 Valence band offsets and band diagrams

Based on Anderson's rule, the VBO at a semiconductor heterojunction is dependent on electron and hole affinities of the materials [43]. In our case, the electron affinities are 3.8 eV and 4.0 eV for AlInP and GaInP, respectively [44]. By taking into account the band gaps (2.3 eV for AlInP and 1.8 eV for GaInP [45]), Anderson's rule suggests a type I heterojunction for the GaInP/AlInP heterostructure, where the band gap of GaInP is within the band gap of AlInP energetically and the VBO is 0.3 eV.

The dependence of VBO on Al and Ga concentration in GaInP/AlInP heterostructures was predicted by density functional theory (DFT) calculation. The VBOs changes from negative values up to 0.56 eV [28]. For the heterostructures with atomic concentration of Ga:In and Al:In equal to ~50:50, the predicted VBO was computed ~ 0.2 eV (a type I of heterojunction).

In order to verify the theoretical predictions, we used Kraut's approach for VBO measurements by XPS [29,46]. In this approach, VBO is derived by measuring the core level peak positions in the heterostructure and the corresponding core level peak positions and VBMs of the references (bulk overlayer and bulk substrate):

$$\Delta E_V = \left(E_{Ga,\ 2p3/2}^{GaInP,\ bulk} - E_{VBM}^{GaInP,bulk}\right) - \left(E_{Al,2p}^{AlInP,bulk} - E_{VBM}^{AlInP,bulk}\right) + \left(E_{Al,2p}^{AlInP/GaInP} - E_{Ga,2p3/2}^{AlInP/GaInP}\right) \quad (1)$$

where $\Delta E_V$ is VBO, $E_{Ga,\ 2p3/2}^{GaInP,\ bulk} - E_{VBM}^{GaInP,bulk}$ is the difference of energy between Ga $2p_{3/2}$ core level and the VBM in 200-nm-thick GaInP(100) sample, $E_{Al,2p}^{AlInP,bulk} - E_{VBM}^{AlInP,bulk}$ is the energy difference between Al 2p and the VBM in the 50-nm-thick AlInP sample, and $E_{Al,2p}^{AlInP/GaInP} - E_{Ga,2p3/2}^{AlInP/GaInP}$ is the difference of Al 2p and Ga $2p_{3/2}$ core level position measured on GaInP/AlInP(100) heterostructure. The measured VBO, $\Delta E_V$, was 0.2 ± 0.04 eV (see Tab. S1 in SI). Therefore, VBO is almost independent on the thickness of AlInP overlayer. The measured value of VBOs agrees very well with the DFT values for $Al_{0.5}In_{0.5}P$ / $Ga_{0.5}In_{0.5}P$ [28].

In Fig. 5, we present schematic band diagram of a) n-GaInP(100) and b) n-AlInP(100) surfaces. The measured VBMs were 1.2 and 1.4 eV for GaInP(100) [see Fig. 2 e), 0 nm sample] and for AlInP(100) (Fig. 2 e), 50 nm sample), correspondingly. The resulting surface BB is $eV_{bb1}$ = ~0.5 eV and $eV_{bb2}$=~0.8 eV was derived by knowing band gap values, $E_{g1}$ and $E_{g2}$.

In Fig. 5 c), band diagram of thin n-GaInP/ n-AlInP(100) heterostructure is shown. Here, thickness of overlayer, $d_L$, is smaller than thickness of space charge layer, $d_{SCL}$. Apart from upward BB, the measured VBO of 0.2 eV and derived conduction band offset (CBO) of 0.3 eV are present. The upward BB is present in the substrate and overlayer due to presence of surface states (Fermi level pinning). A narrow flat-band region in AlInP at interface is suggested based on AR-XPS measurements up to 3-nm-thick overlayer [Fig. 3 b-e]. The band alignment in Fig. 5 c) was measured under the 'dark' conditions, i.e., without illumination by light. During illumination, barriers get reduced by a photovoltage effect [47]. Thus, band flattening and core level peak shifts are expected under laser illumination [48]. We confirm the photovoltage effect in our heterostructure: the core level positions have shifted under illumination of green laser toward the high binding energy side, i.e., upward BB flattening was confirmed (see Fig. SI5 in SI).

In case of ideal n-n type heterojunction [Fig. 5 d] without presence of surface and interface states, downward BB in the substrate and a low upward barrier (typically below 0.3 eV) for the overlayer towards the heterointerface is expected [49]. Therefore, larger thickness of overlayer and passivation of surface electronic states from the heterointerface is needed in order to eliminate the potential barrier at interface. A detailed interface studies of such thick multilayer structures require high photon energies, i.e., HAXPES at synchrotron.

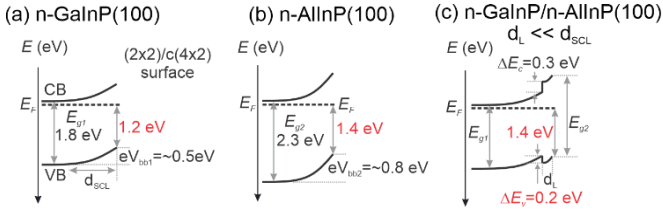

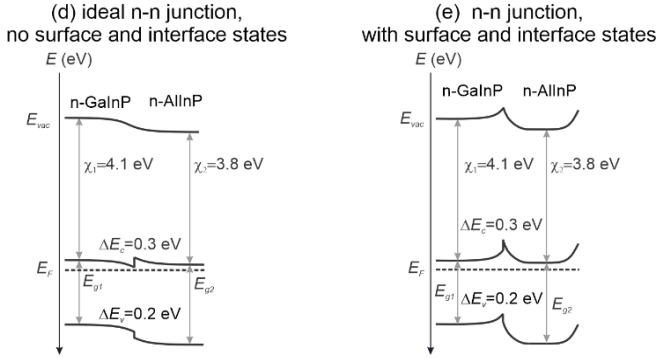

Fig. 5, Measured band alignment at the a) n-GaInP(100), b) n-AlInP(100) surfaces, and (c) thin n-GaInP/n-AlInP(100) heterostructure. Measured values are marked by red. Electron affinities and vacuum levels (Evac) are shown. Theoretical models of band alignment at n-GaInP/n-AlInP junction d) without surface and interface states and e) in a presence of surface and interface states.

In Fig. 5 e), we suggest a general model of band alignment diagram for n-n type heterojunction in a presence of interface and surface states. For this energy diagram the upward band bending of GaInP and AlInP as deduced from the surface studies have been considered together with the known n-type doping concentration, which have been determined by CV measurements. The resulting band energy diagram contains an internal barrier at interface and, despite the barrier height can be reduced by photovoltage effect, it cannot explain the given favorite electron transport across the GaInP/AlInP interface [2–4]. Therefore, we conclude that the interface state or defect density is low in our GaInP/AlInP heterojunction, and we suggest that the band alignment model in Fig. 5 d) is realized in the real devices with passivated surface states.

## 4. Conclusions

Surface and interface structure of MOVPE prepared n-GaInP/AlInP(100) heterostructures were investigated by XPS. We revealed P-rich surface reconstruction on both AlInP and GaInP surfaces. Upward surface band bending on both GaInP and AlInP clean surfaces terminated by (2×2)/c(4×2) surface reconstruction was measured. LEED pattern analysis suggest presence of one-dimensional disorder on both GaInP and AlInP surfaces. We resolved shifted surface-related component in P 2p core levels by AR-XPS, whereas no interface-related core level components have been observed. Furthermore, we derived the experimental VBO of 0.2 eV of GaInP/AlInP heterostructures in agreement with previous theoretical predictions. Finally, our studies predict the formation of the localized mid-gap electronic states on GaInP/AlInP heterostructure surface and open question about origin of these states and their elimination for practical applications.

## Acknowledgments

We are grateful to acknowledge the financial support of German Research Foundation (DFG, PAK981: Project No. HA3096/14-1 and JA859/35-1 and focus program SPP2169: Project No. 423746744). The authors thank A. Muller for his technical support for the series of experiments. OR acknowledges the supported by Operational Program Research, Development and Education financed by European Structural and Investment Funds and the Czech Ministry of Education, Youth and Sports (Project No. SOLID21 - CZ.02.1.01/0.0/0.0/16_019/0000760).

# Supplementary Information

# Band energy diagrams of n-GaInP/n-AlInP(100) surfaces and heterointerfaces studied by X-ray photoelectron spectroscopy

Mohammad Amin Zare Pour[1], Oleksandr Romanyuk[2], Dominik C. Moritz[3], Agnieszka Paszuk[1], Clément Maheu[3], Sahar Shekarabi[1], Kai Daniel Hanke[1], David Ostheimer[1], Thomas Mayer[3], Wolfram Jaegermann[3], Jan P. Hofmann[3], Thomas Hannappel[1,*]

[1]*Institute of Physics, Fundamentals of Energy Materials, Technische Universität Ilmenau, Gustav-Kirchhoff-Str. 5, 98693 Ilmenau, Germany*

[2]*FZU – Institute of Physics of the Czech Academy of Sciences, Cukrovarnicka 10, 16200 Prague, Czech Republic*

[3]*Surface Science Laboratory, Department of Materials and Earth Sciences, Technical University of Darmstadt, Otto-Berndt-Strasse 3, 64287 Darmstadt, Germany*

1. XRD measurements of lattice constants

GaInP(100) and AlInP(100) epilayers with thicknesses of 200 nm were grown in a $H_2$-based, horizontal-flow MOVPE reactor (Aixtron, AIX-200) on n-type GaAs(100) wafer (Si-doped, $2 \times 10^{18}$ atoms/cm$^3$) with 0.1° offcut toward <111> plane. High-resolution X-ray diffraction (XRD) ω/2θ scans of GaAs(004) reflection were measured by Bruker AXS D8 Discover diffractometer. In Fig. SI1, diffraction intensities are shown for a) 200 nm GaInP on GaAs(100) sample, b) 200 nm AlInP on GaAs(100) sample, and c) 200 nm AlInP on 200 nm GaInP on GaAs(100) sample. To avoid oxidation of the AlInP(100) epilayers, the layers were 'capped' with a 20 nm GaAs layer. Diffraction peak maxima lie at 33.05°. Derived In atomic stoichiometry is y=0.51 in $Ga_yIn_{1-y}P$ and x=0.52 in $Al_xIn_{1-x}P$. XRD measurements confirmed the same lattice constants for overlayers and GaAs(100) substrate.

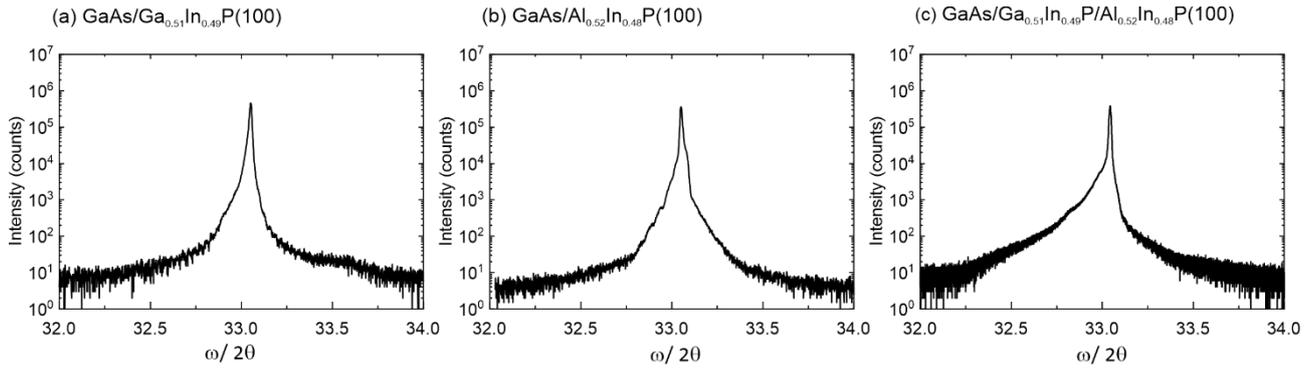

*Fig. SI1 – XRD scans of GaAs(004) reflection from a) 200 nm GaInP on GaAs(100), b) 200 nm AlInP on GaAs(100), and c) 200 nm AlInP on 200 nm GaInP on GaAs(100) heterostructures.*

## 2. ECVP measurements of carrier concentration

The carrier concentration profile was measured by electrochemical capacitance voltage profiling (ECVP, WEP CVP21) method on 10 nm thick AlInP on 100 nm GaInP on GaAs(100) sample. Sample was etched by HCl 1 m. The carrier concentration of ~$1.3 \times 10^{17}$ cm$^{-3}$ was obtained on the surface of the sample. An out-diffusion of Si (n-type dopant) from the GaAs 100 nm thick buffer layer with a doping carrier concentration of ~$4 \times 10^{18}$ cm$^{-3}$ into the GaInP(100) buffer layer is observed. The carrier concentration at the GaInP/AlInP interface is ~$1.4 \times 10^{17}$ cm$^{-3}$ and at the GaAs/GaInP(100) interface is ~ $4 \times 10^{18}$ cm$^{-3}$, which is close to the doping level of the GaAs buffer layer.

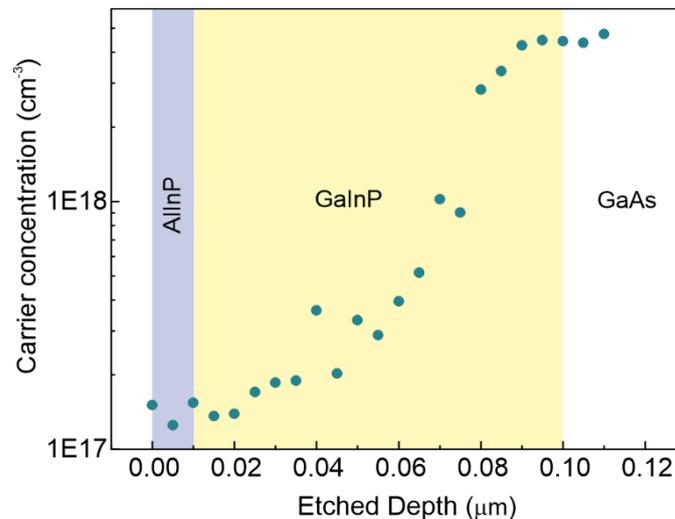

*Fig. SI2 – Carrier concentration profiles obtained on 10-nm-thick AlInP/GaInP sample by CVP.*

## 3. Core level relative shifts and energies

The energy difference between the III-type cation core levels and V-type anion core levels should be constant independently on band bending or doping of the sample. Changes in crystal type (binary vs. ternary semiconductors) or crystal stoichiometry may produce chemical shifts, i.e., binding energy deviation. Therefore, we measured the core level positions in the bulk (binary) and thick heterostructure (ternary) samples.

Fig. SI3 shows XPS core level spectra measured on 100-nm-thick GaAs/GaInP(100), 50-nm-thick GaInP/AlInP(100) samples (UHV-transferred, contamination and oxide free), and on reference oxidized III-V(100) samples. Spectra were calibrated to the position of the P 2p peak maxima (shifted to 129.0 eV). The peak intensities were normalized to unity. Note, quality of the oxidized surfaces is expected to be low than grown clean surfaces and, therefore, the FWHMs of peaks deviate from sample to sample. Below we focus on peak position shifts rather on peak line shapes.

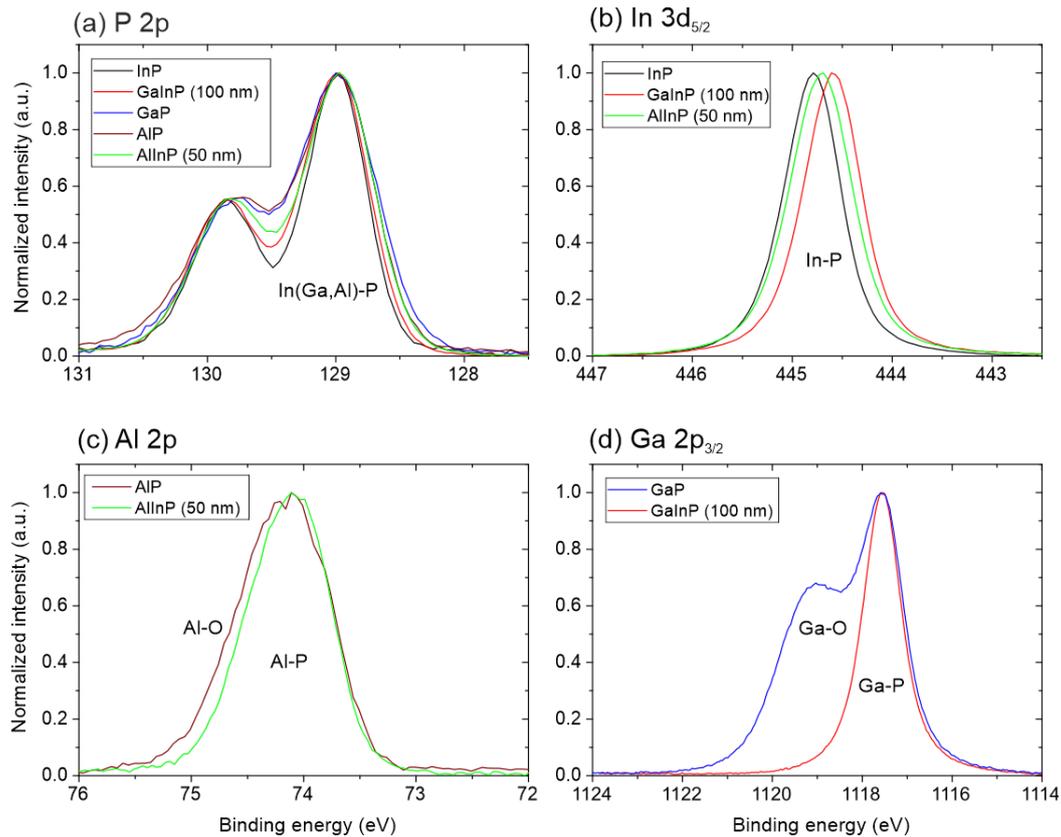

*Fig. SI3 – Calibrated XPS spectra to P 2p position from heterostructure and bulk reference samples. Peak position differences are similar in bulk and heterostructures for P, Al, and Ga. Position of In core levels in respect to P core levels depends on In local environment in a lattice.*

Energy difference between P 2p and Al 2p or Ga $2p_{3/2}$ is the same for bulk and heterostructure sample. There is no dependence on atomic composition of the compound. Chemical shifts are obvious for the In $3d_{5/2}$ peak. Here, binding energy of In-P bond depends on local chemical environment around the In-P bonds.

In Tab. SI1, as-measured (absolute) core level binding energies of bulk reference samples are shown. The core level pair differences in bulk crystals are -54.91 eV [between Al $2p_{3/2}$ – P $2p_{3/2}$ in AlP(100)], 315.76 eV [between In $3d_{5/2}$ – P $2p_{3/2}$ for InP(100)], and 988.56 eV [between Ga $2p_{3/2}$ - P $2p_{3/2}$ for GaP(100)].

*Tab. SI1 – Measured binding energies on oxidized bulk references by XPS.*

|  | Binding energy, eV | | | | |
|---|---|---|---|---|---|
|  | Al $2p_{3/2}$ | Ga $2p_{3/2}$ | In $3d_{5/2}$ | P $2p_{3/2}$ | VBM |
| AlP(100) | 74.04 | - | - | 128.95 | - |
| GaP(100) | - | 1117.14 | - | 128.58 | 0.5 |
| InP(100) | - | - | 444.91 | 129.16 | 1.4 |

*Tab. SI2 – Measured binding energies and VBOs on n-GaInP/n-AlInP(100) heterostructures by XPS.*

| AlInP Thickness, nm | Binding energy, eV | | | | | VBO |
|---|---|---|---|---|---|---|
|  | VBM | Al $2p_{3/2}$ | Ga $2p_{3/2}$ | In $3d_{5/2}$ | P $2p_{3/2}$ |  |
| 0 | 1.2 | 0.00 | 1117.81 | 444.86 | 129.25 | 0.18 |
| 1 | 1.3 | 74.17 | 1117.84 | 444.88 | 129.24 | 0.19 |
| 2 | 1.3 | 74.22 | 1117.88 | 444.92 | 129.24 | 0.16 |
| 3 | 1.4 | 74.28 | 1117.96 | 445.00 | 129.26 | 0.18 |
| 4 | 1.5 | 74.32 | 1117.98 | 445.04 | 129.30 | 0.20 |
| 11 | 1.6 | 74.35 | - | 445.06 | 129.30 | - |
| 32 | 1.4 | 74.20 | - | 444.91 | 129.14 | - |
| 50 | 1.4 | 74.14 | - | 444.85 | 129.11 | - |

In Tab. SI2, as-measured core level binding energies on heterostructure samples are included. In Fig. SI4, we plot relative core level peak position differences in respect to the corresponding difference in bulk crystals (Tab SI1). Change in atomic composition of thin

AlInP overlayers (close to heterointerface, see Fig. 4 in paper) causes binding energy deviation and chemical shifts. For thicker overlayers, energy differences are similar to the differences in in the bulk crystals.

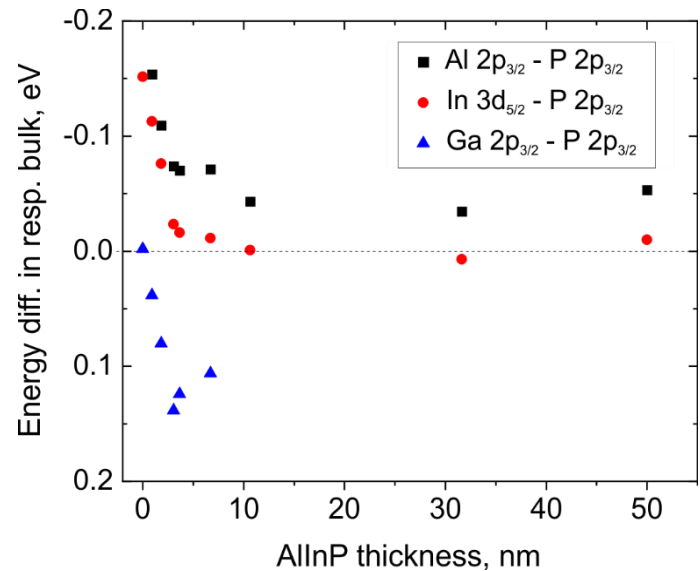

*Fig. SI4 – Relative binding energy difference between the core level pairs measured on heterostructures (see Tab. SI1) in respect to the corresponding difference in the bulk reference crystal (see Tab. SI2).*

### 4. Photovoltaic effect in the n-GaInP/n-AlInP(100) heterostructure

XPS spectra were measured in-situ on 7-nm-thick n-GaInP/n-AlInP(100) heterostructure under dark (without illumination) and under light (with green laser illumination, 532 nm) conditions. Power supply of the laser was turned to 0.5 W and 1.0 W. Laser was set outside of analytical chamber and light was impinged on sample through the window in the chamber. Fig SI4 shows measured spectra. Continuous shift of core levels toward high binding energy side was observed. This effect can be explained by flattening of the bands under illumination conditions: Upward band bending in heterostructure [see Fig. 5 c) in the paper] is getting flat, i.e., distance between Fermi level and core levels is increased. This induces positive shifts of VBMs as well as core level peak maxima.

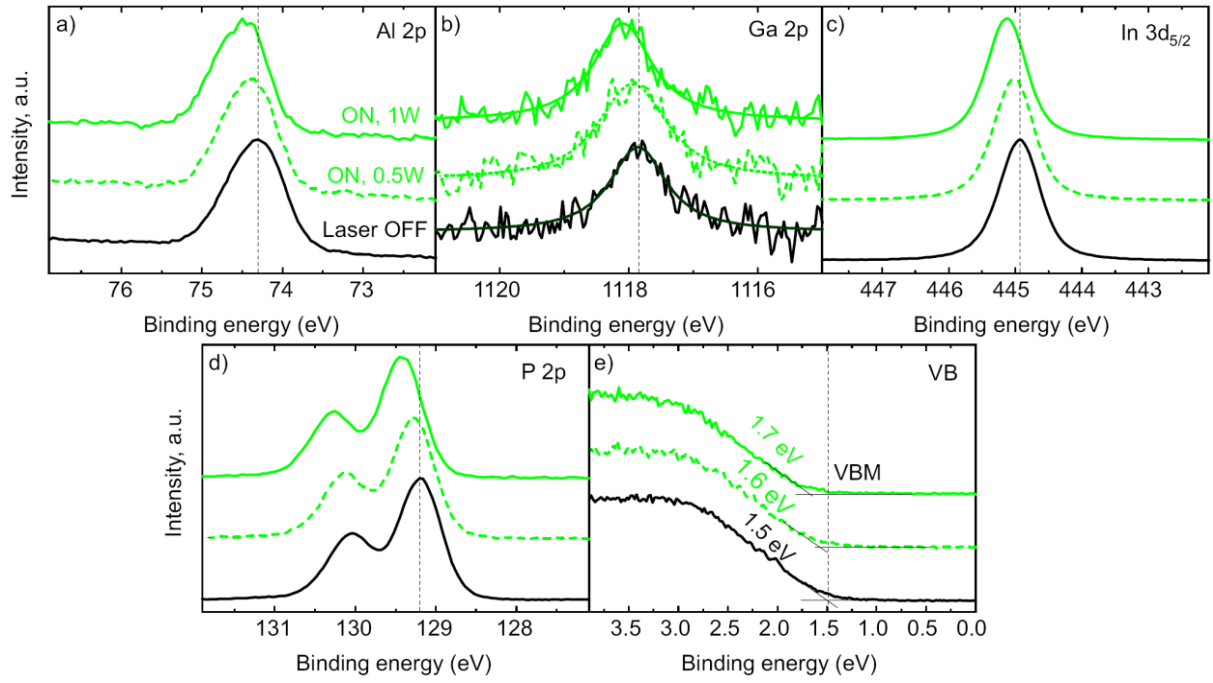

*Fig. SI5 – XPS spectra measured on 7-nm-thick GaInP/AlInP(100) heterostructure under 'dark' condition (laser OFF) and under illuminated condition (green laser ON). Power supply to laser was tuned. A continuous high binding energy shifts for all core levels were observed.*